\documentclass[conference,letterpaper]{IEEEtran}
\usepackage{graphicx,cite,xcolor}
\usepackage{amsmath,amssymb,amsbsy,amscd,amsthm,mathabx}
\usepackage{amsfonts,bm,mathrsfs,bbm}
\usepackage[sans]{dsfont}
\usepackage{mathtools}
\usepackage[nolist]{acronym}
\usepackage{tikz}
\usepackage{soul}
\usepackage{flushend}

\mathtoolsset{showonlyrefs=true}

\IEEEoverridecommandlockouts

\begin{document}
\newtheorem{thm}{Theorem}
\newtheorem{lemma}{Lemma}
\newtheorem{definition}{Definition}
\newtheorem{example}{Example}
\newtheorem{observation}{Remark}
\newtheorem{property}{Property}

\newcommand{\transpose}{{\scriptstyle \mathsf{T}}}
\newcommand{\expect}[1]{\mathbbmss{E}\!\left[#1\right]}
\renewcommand{\Pr}[1]{\mathbbmss{P}\!\left(#1\right)}
\newcommand{\vel}{\vee}
\newcommand{\dec}[1]{\left[#1\right]_{\mathsf{D}}}
\newcommand{\bin}[1]{\left[#1\right]_{\mathsf{B}}}
\newcommand{\hw}{w_{\scriptscriptstyle\mathrm{H}}}
\newcommand{\overeq}[2]{\stackrel{\mathclap{\scriptsize\mbox{#1}}}{#2}}

\newcommand{\syndrome}{\bm{s}}
\newcommand{\testmat}{\bm{A}}
\newcommand{\testmatcol}{\bm{a}}
\newcommand{\defvec}{\bm{x}}
\newcommand{\defvecrv}{\bm{X}}
\newcommand{\testvec}{\bm{t}}
\newcommand{\testvecrv}{\bm{T}}
\newcommand{\defalpha}{\mathcal{X}}
\newcommand{\testalpha}{\mathcal{T}}
\newcommand{\preval}{\delta}
\newcommand{\syndset}[1]{\mathcal{X}_{#1}}
\newcommand{\syndsetunion}{\mathcal{X}}

\newcommand{\pfa}{P_{\scriptscriptstyle\mathrm{FA}}}
\newcommand{\pmd}{P_{\scriptscriptstyle\mathrm{MD}}}
\newcommand{\Lapp}{L^{\scriptscriptstyle\mathrm{APP}}}
\newcommand{\edgeset}[2]{\mathcal{E}_{\scriptscriptstyle #1}^{\scriptscriptstyle (#2)}}
\newcommand{\LRT}[2]{\mathrel{\mathop\gtrless\limits^{#1}_{#2}}}

\renewcommand{\[}{\begin{equation}}
\renewcommand{\]}{\end{equation}}


\title{Optimum Detection of Defective Elements in Non-Adaptive Group Testing}

\author{
	Gianluigi Liva, Enrico Paolini and Marco Chiani 
	\thanks{Gianluigi Liva is with the Institute of Communications and
		Navigation of the German Aerospace Center (DLR), M\"unchner Strasse 20, 82234 We{\ss}ling, Germany.
		Email: gianluigi.liva@dlr.de.}
		\thanks{Enrico Paolini and Marco Chiani are with CNIT, DEI, University of Bologna, via Dell’Universit\`{a} 50, 47522 Cesena (FC), Italy. Email: \{e.paolini,marco.chiani\}@unibo.it.}
}

 \maketitle

\thispagestyle{empty}


\begin{abstract}
We explore the problem of deriving a posteriori probabilities of being defective for the members of a population in the non-adaptive group testing framework. Both noiseless and noisy testing models are addressed. The technique, which relies of a trellis representation of the test constraints, can be applied efficiently to moderate-size populations. The complexity of the approach is discussed and numerical results on the false positive probability vs. false negative probability trade-off are presented.
\end{abstract}

\thispagestyle{empty}
\setcounter{page}{1}

\begin{acronym}
\acro{BP}{belief propagation}
\acro{LDPC}{low-density parity-check}
\acro{MAP}{maximum a posteriori probability}
\acro{r.v.}{random variable}
\acro{i.i.d.}{independent, identically-distributed}
\acro{ML}{maximum likelihood}
\acro{APP}{a posteriori probability}
\acro{COMP}{combinatorial orthogonal matching pursuit}
\acro{LLR}{log-likelihood ratio}
\acro{ROC}{receiver operating characteristic}
\acro{BCH}{Bose–Chaudhuri–Hocquenghem}
\acro{SARS-CoV-2}{severe acute respiratory syndrome coronavirus 2}
\end{acronym}


\section{Introduction}\label{sec:intro}

It is widely acknowledged that large-scale testing and contact tracing play a fundamental role in the containment of \ac{SARS-CoV-2} outbreaks \cite{lavezzo2020suppression}. The task is nevertheless rendered difficult by  limitations in terms of test analysis facilities and trained personnel. Moreover, the cost associated with accurate tests, along with the shortage of required chemical reagents, poses severe challenges in the implementation of a mass testing policy \cite{narayanan2020pooling}. To address the problem, an emerging trend is to use \emph{group testing} techniques \cite{dorfman} as a means of reducing the test analysis time, effort, and costs \cite{mallapaty2020mathematical}. Recently, the use of group testing has been advocated to enable mass testing in the context of the on-going \ac{SARS-CoV-2} pandemic, with experimental campaigns implemented in a few countries \cite{mallapaty2020mathematical}.

Several flavors of group testing have been developed over the years (we point the interested reader to \cite{aldridge2019group} for a through survey). A first, fundamental distinction is between the so-called \emph{adaptive} and \emph{non-adaptive} group testing. In adaptive group testing, the tests are performed in sequence, with pools that are created based on the outcomes of the previous tests. Simple examples of adaptive group testing strategies involve the use of binary search trees \cite{sobel}. On the contrary, in non-adaptive group testing all pools are a-priori set, and tests are carried out in parallel. Both approaches have advantages and shortcomings: adaptive strategies can identify the status of individuals with fewer tests. Nevertheless, considering the time required to carry out each test, a pure adaptive strategy may require more time to determine the status for the each member of the tested population. Non-adaptive schemes require typically more tests to succeed, but they tend to be faster since tests can be performed in parallel. Importantly, non-adaptive group testing algorithms often display a non-trivial trade-off between the false positive and false negative probabilities (i.e., between the \emph{false alarm} and the \emph{miss-detection probabilities}). To combine the advantages of both techniques, while mitigating their limitations, it is sometimes preferable to implement a hybrid approach, where a first screening is performed via a non-adaptive testing step, followed by an adaptive (or even individual) testing step for the population members that are identified as potentially infected. The first step has the role to prune the sample population, delivering to the second step a small fraction of the original set of individuals for the additional testing. Approaches of this kind, which date back to the original work of Dorfman \cite{dorfman}, enable remarkable savings in the number of tests. Several on-going investigations on the use of group testing for \ac{SARS-CoV-2} screening follow this line \cite{schmidt2020detection,heidarzadeh2020two}.

In this paper, we address the problem of efficient \ac{APP} detection of defective elements in a non-adaptive setting. Our work falls along the lines of \cite{sejdinovic-johnson,emad2014,wadayamaISIT}, where belief propagation was used to the detect defective elements. In particular, we investigate the use of a trellis description of the test matrix to enable the use of the forward-backward algorithm \cite{BCJR74}. The technique is reminiscent of the trellis representation of linear block codes based on the parity-check matrix \cite{BCJR74,Wolf_Trellis78}, and allows obtaining \ac{APP} estimates for each element of the population with a complexity that grows exponentially in the number of tests (rather than in the population size). The approach can be applied to small and moderate size test matrices and it may be considered as a building block for more sophisticated group testing strategies \cite{SAFFRON,Karimi19:IrregularGraphs}. It is developed for both noiseless and noisy group testing settings.

The paper is organized as follows. Section \ref{sec:prel} provides the main definitions and the notation used in the rest of the manuscript. Section \ref{sec:Trellis} presents the trellis construction. The application of the forward-backward algorithm (derived in Appendix A) is discussed in Section \ref{sec:BCJR}, along with some numerical examples. Conclusions follow in Section \ref{sec:conc}.

\section{Preliminaries}\label{sec:prel}

We consider a non-adaptive group testing problem where $m$ pooled tests are applied to a population of $n$ elements. The status of the population is described by the \emph{defectivity vector} $\defvec=(x_1,x_2,\ldots,x_n)$ where each element belongs to $\{0,1\}$. For the defectivity vector, we adopt an \ac{i.i.d.} model where each element is defective (i.e., it takes value $1$) with probability $\delta$, where $\delta$ is referred to as the \emph{prevalence}. We denote by $\syndrome=(s_1,\ldots,s_m)$ the \emph{syndrome vector}, where $s_i=0$ if none of the elements of $\defvec$ participating in the $i$th pool is defective while $s_i=1$ if at least one element participating in the pool is defective. The tests are, therefore, non-quantitative. 
The allocation of the population elements to the pools is described by an $m\times n$ binary \emph{test matrix} $\testmat=\left\{a_{i,\ell}\right\}$, where $a_{i,\ell}=1$ if and only if the $\ell$th element of the population participates in the $i$th pool. Compactly, we write
\[
\syndrome:=\defvec \vel \testmat^{\transpose}
\]
where the $\vel$ operator between the vector $\defvec$ and the matrix $\testmat^{\transpose}$ is defined to yield
\[
s_i=\bigvee_{\ell=1}^n \left(x_\ell \wedge a_{i,\ell}\right).
\]
Here, $\vel$ is the inclusive logical disjunction (``or'') and $\wedge$ is the logical conjunction (``and'').
We consider two models for the tests. In a first (noiseless) model, the \emph{test vector} $\testvec$ is equal to the syndrome,  $\testvec=\syndrome$,
i.e., tests are error-free. In a second model, we observe a noisy version of the syndrome, yielding a test vector that is only statistically dependent on the syndrome according to a generic distribution $Q(\testvec|\syndrome)$. We further assume the test vector to take values in $\{0,1\}^m$.
The random vectors associated with $\defvec$ and $\testvec$ are indicated as $\defvecrv$ and $\testvecrv$, respectively.
We denote the set of defectivity vectors compatible with a syndrome $\syndrome$ as
\[
\syndset{\syndrome}:=\left\{\defvec | \defvec \vel \testmat^{\transpose}=\syndrome\right\}.
\]
The decision taken on the status of the elements is $\hat{\defvec}$ (and $\hat{\defvecrv}$ is the corresponding random vector). The false-alarm probability is 
\[
\pfa:=\frac{1}{n}\sum_{\ell=1}^{n}\Pr{\hat{X}_\ell=1 | X_\ell=0}
\]
and the miss-detection probability is 
\[
\pmd:=\frac{1}{n}\sum_{\ell=1}^{n}\Pr{\hat{X}_\ell=0 | X_\ell=1}.
\] 
In the following, $\log$ is the natural logarithm, and $\hw(\defvec)$ is the Hamming weight of the  vector $\defvec$.

\section{Trellis Diagram Construction based on the Test Matrix}\label{sec:Trellis}

In this section, we illustrate how the sets of defectivity vectors $\syndset{\syndrome}$ can be compactly represented through a trellis diagram with $n$ sections and at most $2^m$ states per section. The trellis construction follows the footsteps of the construction introduced in \cite{BCJR74,Wolf_Trellis78} to represent a linear block code based on the code parity-check matrix.

We denote by $S_\ell$ the state at \emph{depth} $\ell$, where the state can take value in $\left\{0,1,\ldots,2^m-1\right\}$. We further introduce the \emph{partial syndrome vector} at depth $\ell$ as $\syndrome_\ell$. Observe that the syndrome can be obtained as
\[
\syndrome=\bigvee_{\ell=1}^n \left(x_\ell \wedge \testmatcol_\ell^\transpose\right)
\]
where $\testmatcol_\ell$ is the $\ell$th column of the test matrix and the $\wedge$-operation has to be intended as element-wise. Owing to the associativity of the $\vee$ operator, we can obtain $\syndrome=\syndrome_n$ following the recursion
\[
\syndrome_\ell=\syndrome_{\ell-1} \vee \left(x_\ell \wedge \testmatcol_\ell^\transpose\right)
\]
for $\ell=1,\ldots,n$, and where $\syndrome_0 := (0,0,\ldots,0)$.
Following this observation, we associate to each possible partial syndrome $\syndrome_\ell$ the state at depth $\ell$ with index equal to the decimal representation of the syndrome. Specifically, to a syndrome $\syndrome=(s_1,s_2, \ldots, s_m)$ we associate the state index $\dec{\syndrome}=\sum_{i=1}^m s_i 2^{i-1}$. Similarly, we retrieve the syndrome associated with a state $S$ as the binary expansion of the state index through the operator $\bin{S}$, i.e., $\syndrome=\bin{\dec{\syndrome}}$.
The trellis construction proceeds as follow. At depth $0$, the trellis admits only state $0$. At depth $1$, two states $\dec{x_1 \wedge \testmatcol_1}$ for $x_1\in\{0,1\}$ are allowed: it is easy to check that the first state is (again) state $0$, and that the second state has index $\dec{\testmatcol_1}$. We then connect state $0$ at depth $0$ to state $0$ at depth $1$ through a $0$-labeled edge (i.e., associated to $x_1=0$), and to state $\dec{\testmatcol_1}$ through a $1$-labeled edge (i.e., associated to $x_1=1$). The construction proceeds recursively: For each admitted state $S_{\ell-1}$ at depth $\ell-1$, we draw an $x_\ell$-labeled edge connecting to state $S_{\ell}$ if and only if  
\[
S_{\ell}=\dec{x_\ell \vee \bin{S_{\ell-1}}}.
\]
 The construction proceeds recursively until $\ell=n$. We refer to the trellis obtained by following this procedure as the \emph{complete} trellis.

\medskip

\begin{figure}[t]
	\begin{center}
		\includegraphics[width=\columnwidth]{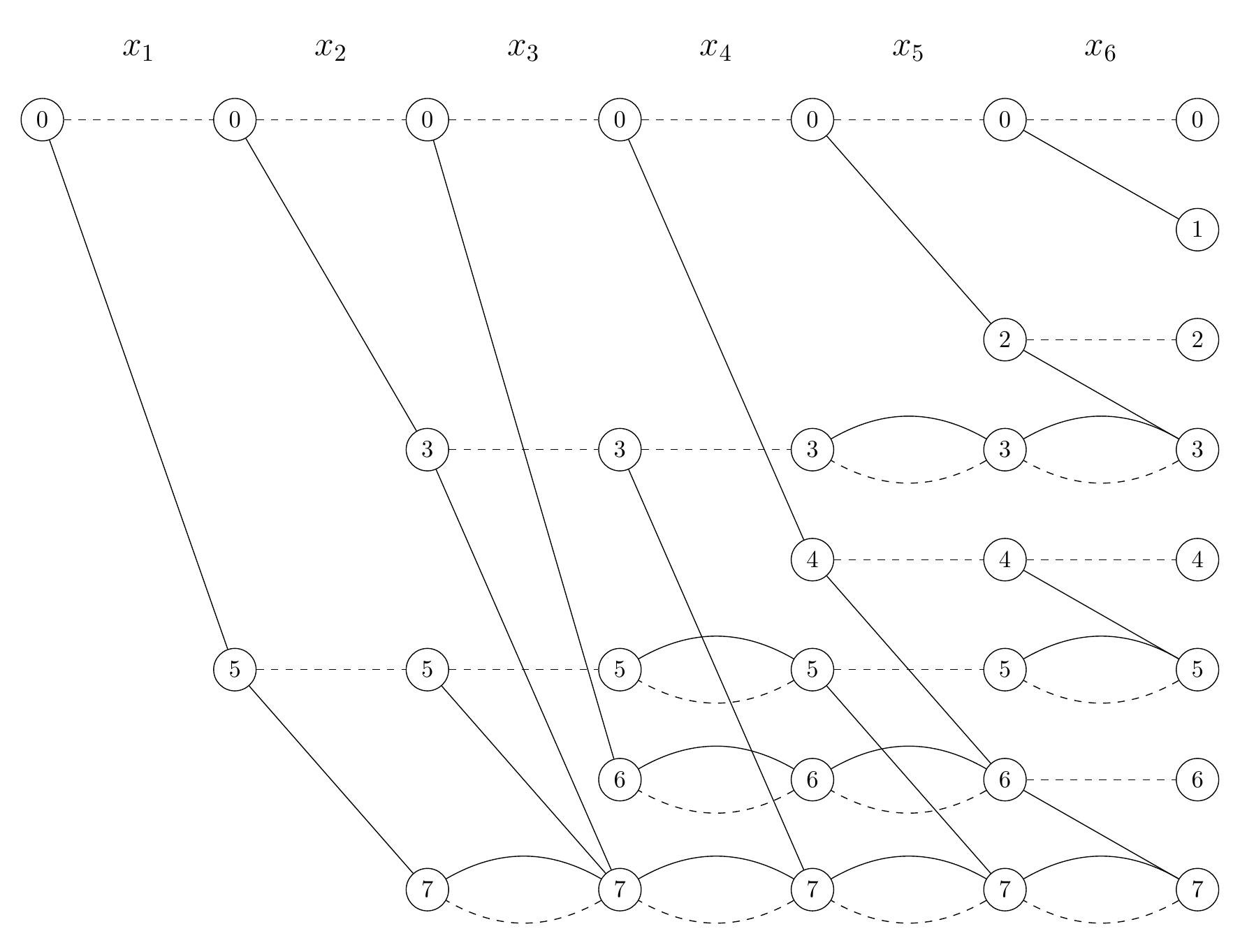}
	\end{center}
	\caption{Complete trellis diagram for the test matrix described in Example \ref{example1}. Dashed lines are used to denote $0$-labeled edges, whereas solid lines are used for $1$-labeled edges.
	}\label{fig:trellis47}
\end{figure}

\begin{example}\label{example1}
	Consider a setting where $n=6$ elements are pooled according to the test matrix
	\[
	\testmat=
	\begin{pmatrix}
		1 & 1 & 0 & 1 & 0 & 0 \\
		0 & 1 & 1 & 0 & 1 & 0 \\
		1 & 0 & 1 & 0 & 0 & 1 
	\end{pmatrix}.
	\]
	The corresponding trellis diagram is depicted in Figure \ref{fig:trellis47}.  
\end{example}

\medskip

Note that all paths reaching the final state $\dec{\syndrome}$ correspond to the defectivity vectors in $\syndset{\syndrome}$. Note also that the trellis diagram may present parallel edges between two states.
The trellis diagram can be used to efficiently obtain the \ac{APP} $\Pr{x_\ell|\testvec}$ for each element in $\defvec$ via the forward-backward algorithm \cite{BCJR74}, as it will be illustrated in Section \ref{sec:BCJR}. Before proceeding, we will highlight some features of the trellis representation that are important in the noiseless group testing setting.

\medskip

\begin{figure}[t]
	\begin{center}
		\includegraphics[width=\columnwidth]{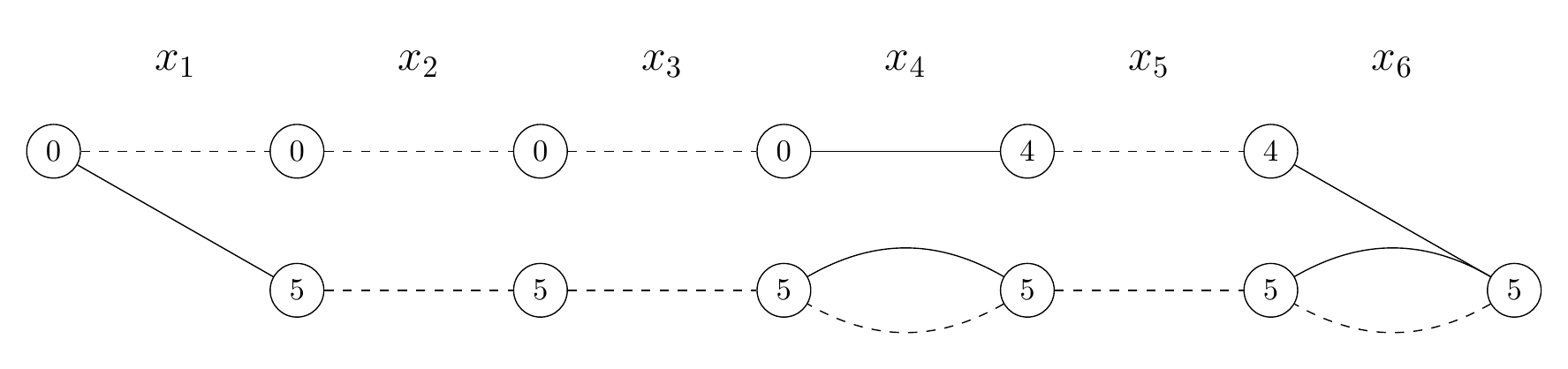}
	\end{center}
	\caption{Expurgated trellis diagram for the test matrix described in Example~\ref{example1}, in the noiseless group testing setting, for the case where $\testvec=(1,0,1)$. Dashed lines are used to denote $0$-labeled edges, whereas solid lines are used for $1$-labeled edges.
	}\label{fig:trellis47comp}
\end{figure}

\begin{observation}\label{observation_states}
	In a noiseless group testing setting (i.e., where $\testvec=\syndrome$), upon observing the test vector $\testvec$ the trellis diagram can be expurgated by removing all paths that do not terminate at the state $\dec{\testvec}$. This can be done without incurring in any loss of information. The paths removal leads to an \emph{expurgated} trellis diagram with a (possibly) reduced number of states. The paths contained in the new trellis correspond to defectivity vectors compatible with the syndrome $\syndrome$, i.e., all vectors in $\syndset{\syndrome}$. Following Example \ref{example1}, Figure \ref{fig:trellis47comp} reports the trellis associated to a final state $\dec{(1,0,1)}=5$. In a noiseless group testing setting, following \cite{Wolf_Trellis78}, we refer to the trellis obtained by removing all paths that do not yield the observed syndrome as the \emph{expurgated} trellis.
\end{observation}

	By visual inspection of the expurgated trellis of Figure \ref{fig:trellis47comp}, we see that the second, third, and fifth trellis sections contain only $0$-labeled edges, i.e., $x_2=x_3=x_5=0$ with certainty.
This fact is not surprising, since, whenever a given test evaluates at $0$, the elements in $\defvec$ participating in the test can be surely marked as non-defective as foreseen, for example, by the \ac{COMP} algorithm \cite{kautz,aldridge2019group}.
In light of this, the following property holds.

\medskip

\begin{property}\label{prop:trellis}
	Denote by $m_0$ the number of non-zero tests in $\testvec$ (i.e., $m_0=\hw(\testvec)$), and by $n_0$ the number of elements in $\defvec$ which participate only in pools resulting in a non-zero test. Then,
	in a noiseless group testing setting (i.e., where $\testvec=\syndrome$), upon observing the test vector $\testvec$ the trellis diagram can be reduced to a trellis with $n_0$ sections and at most $2^{m_0}$ states per section. 
\end{property}

\medskip

\begin{figure}[t]
	\begin{center}
		\includegraphics[width=0.55\columnwidth]{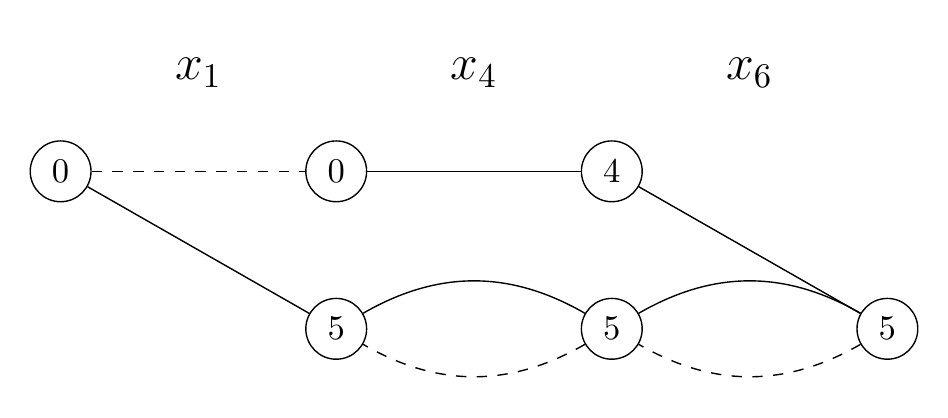}
	\end{center}
	\caption{Reduced trellis diagram for the test matrix described in Example \ref{example1}, in the noiseless group testing setting, for the case where $\testvec=(1,0,1)$. Dashed lines are used to denote $0$-labeled edges, whereas solid lines are used for $1$-labeled edges.
	}\label{fig:trellis47comp2}
\end{figure}

We refer to the trellis following from Property \ref{prop:trellis} as the \emph{reduced} trellis associated with the test vector $\testvec$. Figure \ref{fig:trellis47comp2} provides the reduced trellis for $\testvec=(1,0,1)$, for the test matrix of Example \ref{example1}. Note that, in a noiseless group testing setting, the possibility of describing the whole set of defectivity vectors with a reduced trellis possessing at most $2^{m_0}$ states per section enables dramatic savings on the average complexity of the detection algorithm provided in the next section.

\section{Detection via Forward-Backward Algorithm}\label{sec:BCJR}

Let us consider the general case of a noisy group testing setting as described in Section \ref{sec:prel}. We are interested in evaluating the logarithmic \ac{APP} ratio
\[\label{eq:Lapp}
\Lapp_\ell := \log \left[\frac{\Pr{X_\ell=0 | \testvecrv=\testvec}}{\Pr{X_\ell=1 | \testvecrv=\testvec}}\right].
\]
By means of the complete trellis representation introduced in Section \ref{sec:Trellis}, \eqref{eq:Lapp} can be computed efficiently via the forward-backward algorithm \cite{BCJR74} as
\[
	\begin{aligned}
	\Lapp_\ell &= \log \displaystyle \sum_{(s',s)\in \edgeset{\ell}{0}} \alpha_{\ell-1}(s')\gamma_\ell(s',s)\beta_\ell(s) \\&-\log \sum_{(s',s)\in \edgeset{\ell}{1}} \alpha_{\ell-1}(s')\gamma_\ell(s',s)\beta_\ell(s).\label{eq:FB_final}
	\end{aligned}
\]
In \eqref{eq:FB_final}, $\edgeset{\ell}{x}$ is the set of $x$-labeled edges in section $\ell$, and $(s',s)$ denotes an edge connecting state $s'$ at depth $\ell-1$ with state $s$ at depth $\ell$. Moreover, the \emph{forward} metric at state $s$ and depth $\ell$ can be recursively computed as
\[
\alpha_\ell(s)=\sum_{s'} \alpha_{\ell-1}(s')\gamma_\ell(s',s)\label{eq:forward}
\]
and the \emph{backward} metric at state $s'$ and depth $\ell$ can be obtained as
\[
\beta_{\ell}(s')=\sum_{s} \beta_{\ell+1}(s)\gamma_{\ell+1}(s',s)\label{eq:backward}
\]
with
\[
\gamma_\ell(s',s)=\left\{\begin{array}{ll}
	1-\preval & \quad \textrm{if}\quad (s',s) \in \edgeset{\ell}{0}\\
	\preval & \quad \textrm{if}\quad (s',s) \in \edgeset{\ell}{1}.
	\end{array}\right.
\]
The initial condition for the recursion \eqref{eq:forward} is $\alpha_0(0)=1$ and $\alpha_0(s')=0$ for $s'=1,\dots,2^m-1$, whereas for the backward recursion \eqref{eq:backward} it is  $\beta_n(s)=Q\left(\testvec\,|\,\bin{s}\right)$ for $s=0,\dots,2^m-1$. For sake of completeness, the derivation of \eqref{eq:FB_final}, as well as of \eqref{eq:forward}, \eqref{eq:backward}, is provided in the Appendix.

\medskip

\begin{observation}
For the special case of a noiseless group testing setting, the likelihood $Q\left(\testvec\,|\, \syndrome \right)$ takes value $1$ for $\testvec=\syndrome$, and it is $0$ otherwise. It follows that the forward-backward algorithm can be run on the expurgated (or on the reduced) trellis associated with the syndrome $\syndrome$, by initializing the backward metric to $\beta_n\left(\dec{\syndrome}\right)=1$. Note also that  \eqref{eq:Lapp} can be obtained, in the noiseless setting, by observing that  $\Pr{X_\ell=0 | \testvecrv=\testvec}$ and $\Pr{X_\ell=1 | \testvecrv=\testvec}$ are
\[
\Pr{X_\ell=0 | \testvecrv=\testvec}=\sum_{\substack{\defvec\in\syndset{\testvec}\\x_\ell=0}} \preval^{\hw(\defvec)} (1-\preval)^{n-\hw(\defvec)} \label{eq:brute0}
\]
and
\[
\Pr{X_\ell=1 | \testvecrv=\testvec}=\sum_{\substack{\defvec\in\syndset{\testvec}\\x_\ell=1}} \preval^{\hw(\defvec)} (1-\preval)^{n-\hw(\defvec)}. \label{eq:brute1}
\]
In this case, the forward-backward algorithm can be seen as an efficient way to attack the enumeration problem entailed by \eqref{eq:brute0}, \eqref{eq:brute1}.
\end{observation}

\medskip

A decision about each element in $\defvec$ can be obtained by applying a threshold test to \eqref{eq:Lapp}, i.e.
\[
\Lapp_\ell \LRT{\hat{x}_\ell=0}{\hat{x}_\ell=1} \Lambda \label{eq:appthreshold}
\]
or, by recasting \eqref{eq:appthreshold} as a \ac{LLR} test, as
\[
L_\ell  \LRT{\hat{x}_\ell=0}{\hat{x}_\ell=1} \Lambda'  \label{eq:LLRthreshold}
\]
where 
\begin{align}
L_\ell := \log \left[\frac{\Pr{ \testvecrv=\testvec | X_\ell=0 }}{\Pr{\testvecrv=\testvec | X_\ell=1}}\right]
\end{align}
and
\[
\Lambda'=\Lambda-\log \left[\frac{1-\delta}{\delta}\right].
\]
The test \eqref{eq:LLRthreshold} is optimal in the Neyman-Pearson sense. Moreover, for fixed $\delta$ and a given noise model $Q(\testvec|\syndrome)$, the forward-backward algorithm is deterministic, since it associates to each test vector $\testvec$ a fixed logarithmic \ac{APP} ratio vector $(\Lapp_1,\Lapp_2,\ldots,\Lapp_n)$. It follows that, for a given threshold $\Lambda$, the final decision $\hat{\defvec}$ is fixed and only a discrete set of pairs $(\pmd,\pfa)$ can be achieved, with the operating points linearly interpolating two pairs $(\pmd(\Lambda_1),\pfa(\Lambda_1))$ and $(\pmd(\Lambda_2),\pfa(\Lambda_2))$ achievable through randomized tests. In the noiseless setting, by fixing the threshold $\Lambda$ to a large value, we recover the  \ac{COMP} algorithm \cite{kautz,aldridge2019group}.

\begin{figure}[t]
	\begin{center}
		\includegraphics[width=0.95\columnwidth]{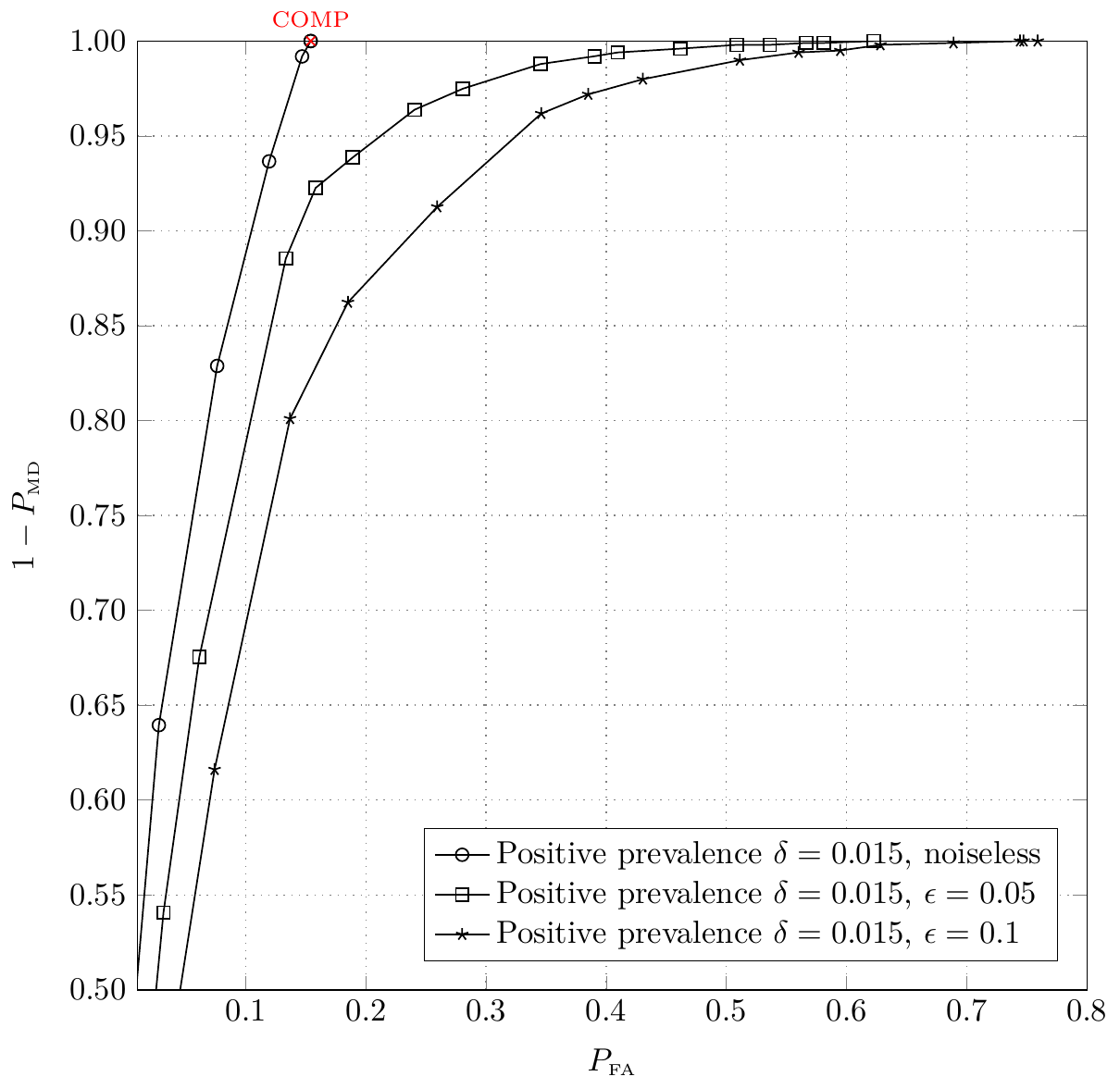}
	\end{center}
	\caption{\ac{ROC} curve for a $7 \times 64$ test matrix $\testmat$, where $\testmat$ is given by the parity-check matrix of a $(64,57)$ extended \ac{BCH} code in cyclic form, where the Hamming weight of each row is $32$.}\label{fig:ROC6457}
\end{figure}

Borrowing from the jargon of detection theory, the \ac{ROC} curves (displaying the probability of successful detection $1-\pmd$ vs. the probability of false alarm $\pfa$ as the threshold $\Lambda$ varies) for a $7\times 64$ test matrix is given in Figure \ref{fig:ROC6457}. The curves have been obtained via Monte Carlo simulations. The test matrix is based on the parity-check matrix of a $(64,57)$ extended \ac{BCH} code in cyclic form, where the Hamming weight of each row is $32$. The \ac{ROC} curves are provided for a prevalence $\delta=0.015$ and for both noiseless and noisy settings. In the noisy case, the noise model mimics the observation of the syndrome through a binary symmetric channel with crossover probability $\epsilon$, i.e., 
\[
Q(\testvec|\syndrome)=\prod_{i=1}^m Q(t_i|s_i)
\]
with $Q(0|0)=Q(1|1)=1-\epsilon$ and $Q(1|0)=Q(0|1)=\epsilon$. In particular, two crossover probabilities are considered, $\epsilon=0.05$ and $\epsilon=0.1$.  In the noiseless setting, by setting $\Lambda$ to a large value we obtain the working point of the \ac{COMP} algorithm, characterized by a zero miss-detection probability. The impact of imperfect tests is remarkable already for a test accuracy of $95\%$ ($\epsilon=0.05$), where to achieve a $98\%$ success rate in the detection the rate of false alarms has to be as high as $30\%$.

\begin{figure}[t]
	\begin{center}
		\includegraphics[width=0.95\columnwidth]{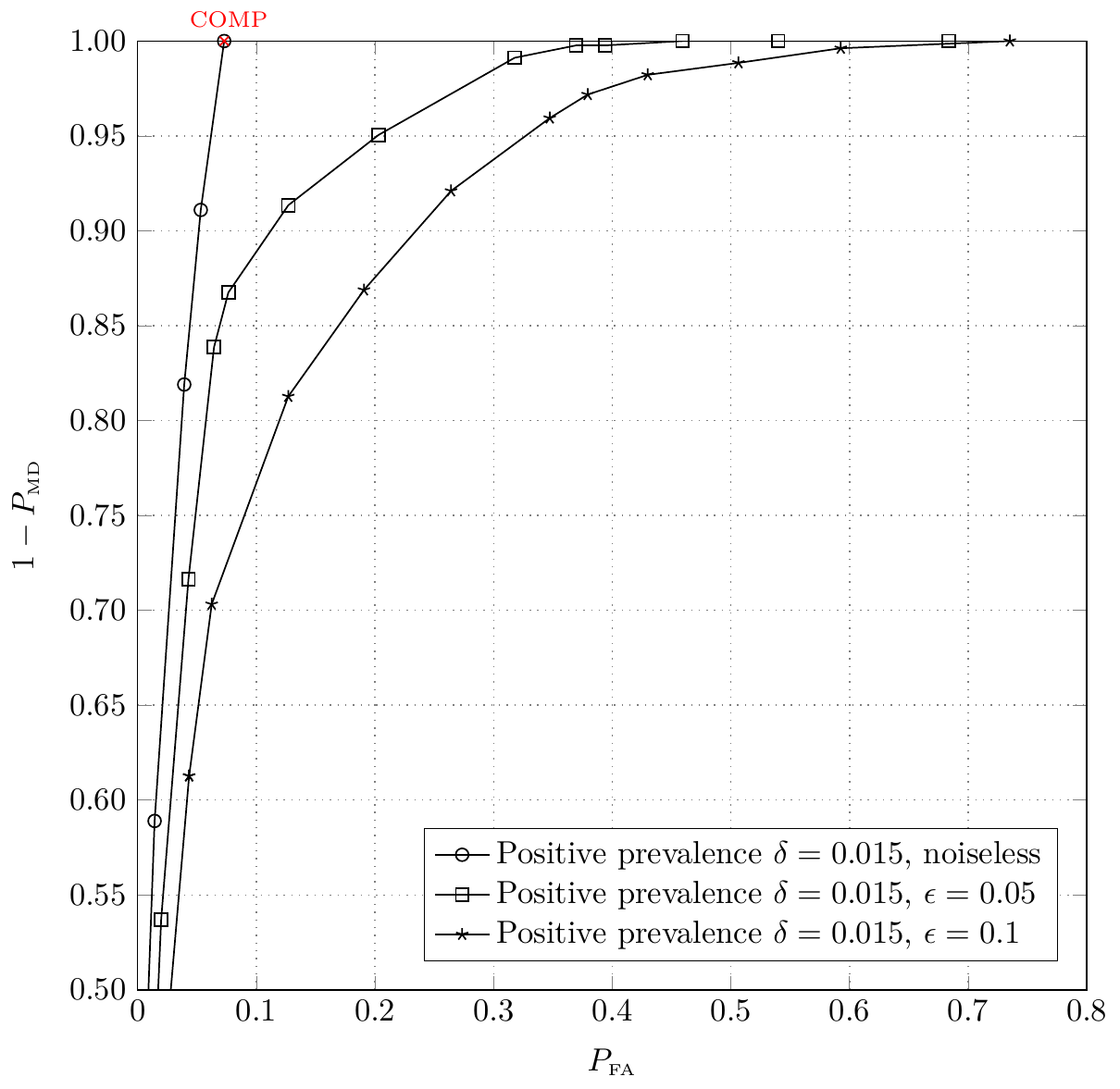}
	\end{center}
	\caption{\ac{ROC} curve for a $9 \times 84$ test matrix $\testmat$, where $\testmat$ is the incidence matrix of an order-$9$, $3$-uniform complete hypergraph.
	}\label{fig:ROCK3_84}
\end{figure}

Figure \ref{fig:ROCK3_84} reports the \ac{ROC} curves for the same conditions considered in the previous example, for the case where the $9 \times 84$ test matrix is given by the incidence matrix of an order-$9$, $3$-uniform complete hypergraph (i.e., each column has Hamming weight $3$ and the the matrix $\testmat$ is composed by the set of all possible weight-$3$ columns).  

An open question relates to the test matrix design criteria that, for given matrix dimensions, provide the best miss-detection vs. false-alarm probability trade-off under the forward-backward detection algorithm.

\section{Conclusions}\label{sec:conc}

 In this paper, we addressed the problem of deriving a posteriori probabilities of being defective for the members of a population in the non-adaptive, non-quantitative group testing framework, both in the noiseless and noisy settings. The approach relies on a trellis representation of the test constraints and it can be applied efficiently to testing matrices involving a moderate number of tests. The peculiarities of the technique, when applied to the noiseless setting, are discussed, emphasizing the implications on the complexity of the algorithm. Numerical results on the false positive probability vs. false negative probability trade-off are presented. The approach may be applied also to the scheme of \cite{SAFFRON,Karimi19:IrregularGraphs}, where the algorithm can be employed at the level of the signature matrices. An open research direction is to find (classes of) test matrices capable of providing the best miss-detection vs. false-alarm probability trade-off under the forward-backward detection algorithm.


\appendices

\section{Derivation of the Forward-Backward Algorithm}

By Bayes' rule, rewrite \eqref{eq:Lapp} as
\[
\begin{aligned}
	\Lapp_\ell &= \log \displaystyle \sum_{(s',s)\in \edgeset{\ell}{0}} \Pr{S_{\ell-1}=s',S_{\ell}=s,\testvecrv=\testvec} \\&-\log \sum_{(s',s)\in \edgeset{\ell}{1}} \Pr{S_{\ell-1}=s',S_{\ell}=s,\testvecrv=\testvec}
\end{aligned}
\]
and observe that
\begin{align}
\Pr{S_{\ell-1}=s',S_{\ell}=s,\testvecrv=\testvec}=\hspace{-4cm}&\\[1mm]
&\overeq{(a)}{=}\Pr{\testvecrv=\testvec \, | \,S_{\ell-1}=s',S_{\ell}=s}\Pr{S_{\ell-1}=s',S_{\ell}=s}\\[-2.5ex]
&\overeq{(b)}{=}\Pr{\testvecrv=\testvec \, | \,S_{\ell}=s}\Pr{S_{\ell-1}=s',S_{\ell}=s}\\
&\overeq{(c)}{=}\Pr{\testvecrv=\testvec \, | \,S_{\ell}=s}\Pr{S_{\ell}=s|S_{\ell-1}=s'}\Pr{S_{\ell-1}=s'}
\end{align}
where (a) follows from Bayes' rule, and (b) is due to the fact that the final state depends on the state at depth $\ell-1$ through the state at depth $\ell$. Furthermore, (c) is obtained again by application of Bayes' rule. We introduce the shorthand 
\begin{align}
	\alpha_{\ell-1}(s')&:=\Pr{S_{\ell-1}=s'}\\
	\beta_{\ell}(s)&:=\Pr{\testvecrv=\testvec \, | \,S_{\ell}=s}\\
	\gamma_{\ell}(s',s)&:=\Pr{S_{\ell}=s|S_{\ell-1}=s'}.
\end{align}
Observe that
\[
\Pr{S_{\ell}=s|S_{\ell-1}=s'}=\left\{\begin{array}{ll}
	1-\preval & \quad \textrm{if}\quad (s',s) \in \edgeset{\ell}{0}\\
	\preval & \quad \textrm{if}\quad (s',s) \in \edgeset{\ell}{1}.
\end{array}\right.
\]
and that
\begin{align}
\alpha_{\ell}(s)& = \Pr{S_{\ell}=s}\\
& \overeq{(a)}{=} \sum_{s'}\Pr{S_{\ell-1}=s',S_{\ell}=s}\\
& \overeq{(b)}{=} \sum_{s'}\Pr{S_{\ell}=s| S_{\ell-1}=s'}\Pr{S_{\ell-1}=s'}\\
& = \sum_{s'} \alpha_{\ell-1}(s')\gamma_{\ell}(s',s)
\end{align}
where (a) is due to the total probability theorem, and (b) is due to Bayes' rule. Similarly,
\begin{align}
\beta_\ell(s')\!	& 	=\!	\Pr{\testvecrv=\testvec  | S_{\ell}=s'} \\
				&	\overeq{(a)}{=}\! \sum_s \Pr{\testvecrv=\testvec ,  S_{\ell+1}=s| S_{\ell}=s'} \\
				&	\overeq{(b)}{=}\! \sum_s \Pr{\testvecrv=\testvec  | S_{\ell}=s', S_{\ell+1}=s }\Pr{S_{\ell+1}=s | S_{\ell}=s'} \\[-2.5ex]
				&	\overeq{(c)}{=}\! \sum_s \Pr{\testvecrv=\testvec  | S_{\ell+1}=s }\Pr{S_{\ell+1}=s | S_{\ell}=s'} \\
				&	=\! \sum_s \beta_{\ell+1}(s)\gamma_{\ell+1}(s',s) \\
\end{align}
where (a) is again due to the total probability theorem, (b) from Bayes's rule, and (c) by observing that the final state depends on the state at depth $\ell$ through the state at depth $\ell+1$.


\begin{thebibliography}{10}
	\providecommand{\url}[1]{#1}
	\csname url@samestyle\endcsname
	\providecommand{\newblock}{\relax}
	\providecommand{\bibinfo}[2]{#2}
	\providecommand{\BIBentrySTDinterwordspacing}{\spaceskip=0pt\relax}
	\providecommand{\BIBentryALTinterwordstretchfactor}{4}
	\providecommand{\BIBentryALTinterwordspacing}{\spaceskip=\fontdimen2\font plus
		\BIBentryALTinterwordstretchfactor\fontdimen3\font minus
		\fontdimen4\font\relax}
	\providecommand{\BIBforeignlanguage}[2]{{%
			\expandafter\ifx\csname l@#1\endcsname\relax
			\typeout{** WARNING: IEEEtran.bst: No hyphenation pattern has been}%
			\typeout{** loaded for the language `#1'. Using the pattern for}%
			\typeout{** the default language instead.}%
			\else
			\language=\csname l@#1\endcsname
			\fi
			#2}}
	\providecommand{\BIBdecl}{\relax}
	\BIBdecl
	
	\bibitem{lavezzo2020suppression}
	E.~Lavezzo \emph{et~al.}, ``{Suppression of a SARS-CoV-2 outbreak in the
		Italian municipality of Vo’},'' \emph{Nature}, vol. 584, no. 7821, pp.
	425--429, 2020.
	
	\bibitem{narayanan2020pooling}
	K.~Narayanan \emph{et~al.}, ``{Pooling RT-PCR or NGS samples has the potential
		to cost-effectively generate estimates of COVID-19 prevalence in resource
		limited environments},'' \emph{medRxiv}, 2020.
	
	\bibitem{dorfman}
	R.~Dorfman, ``The detection of defective members of large populations,''
	\emph{The Annals of Mathematical Statistics}, vol.~14, no.~4, pp. 436--440,
	1943.
	
	\bibitem{mallapaty2020mathematical}
	S.~Mallapaty, ``The mathematical strategy that could transform coronavirus
	testing,'' \emph{Nature}, vol. 583, no. 7817, pp. 504--505, 2020.
	
	\bibitem{aldridge2019group}
	M.~Aldridge, O.~Johnson, and J.~Scarlett, ``Group testing: An information
	theory perspective,'' \emph{Foundations and Trends in Communications and
		Information Theory}, 2019.
	
	\bibitem{sobel}
	M.~Sobel and P.~A. Groll, ``Group testing to eliminate efficiently all
	defectives in a binomial sample,'' \emph{Bell Labs Technical Journal},
	vol.~38, no.~5, pp. 1179--1252, 1959.
	
	\bibitem{schmidt2020detection}
	M.~Schmidt, E.~Seifried, S.~Ciesek, and A.~Berger, ``Detection of {SARS-CoV-2}
	in a plurality of biological samples,'' Nov.~5 2020, {US Patent App.
		16/932,487}.
	
	\bibitem{heidarzadeh2020two}
	A.~Heidarzadeh and K.~R. Narayanan, ``{Two-Stage Adaptive Pooling with RT-qPCR
		for COVID-19 Screening},'' \emph{medRxiv}, 2020.
	
	\bibitem{sejdinovic-johnson}
	D.~Sejdinovic and O.~T. Johnson, ``Note on noisy group testing: {A}symptotic
	bounds and belief propagation reconstruction,'' in \emph{Proc. 48th Annual
		Allerton Conference on Communication, Control, and Computing}, Sep. 2010, pp.
	998--1003.
	
	\bibitem{emad2014}
	A.~Emad and O.~Milenkovic, ``Semiquantitative group testing,'' \emph{{IEEE}
		Trans. Inf. Theory}, vol.~60, no.~8, pp. 4614--4636, 2014.
	
	\bibitem{wadayamaISIT}
	T.~{Wadayama}, T.~{Izumi}, and K.~{Mimura}, ``Bitwise {MAP} estimation for
	group testing based on holographic transformation,'' in \emph{Proc. IEEE
		International Symposium on Information Theory (ISIT)}, Jun. 2015.
	
	\bibitem{BCJR74}
	L.~{Bahl}, J.~{Cocke}, F.~{Jelinek}, and J.~{Raviv}, ``Optimal decoding of
	linear codes for minimizing symbol error rate (corresp.),'' \emph{{IEEE}
		Trans. Inf. Theory}, vol.~20, no.~2, pp. 284--287, Mar. 1974.
	
	\bibitem{Wolf_Trellis78}
	J.~{Wolf}, ``Efficient maximum likelihood decoding of linear block codes using
	a trellis,'' \emph{{IEEE} Trans. Inf. Theory}, vol.~24, no.~1, pp. 76--80,
	Jan. 1978.
	
	\bibitem{SAFFRON}
	K.~{Lee}, K.~{Chandrasekher}, R.~{Pedarsani}, and K.~{Ramchandran},
	``{SAFFRON:} a fast, efficient, and robust framework for group testing based
	on sparse-graph codes,'' \emph{{IEEE} Trans. Signal Process.}, vol.~67,
	no.~17, pp. 4649--4664, Sep. 2019.
	
	\bibitem{Karimi19:IrregularGraphs}
	E.~{Karimi}, F.~{Kazemi}, A.~{Heidarzadeh}, K.~R. {Narayanan}, and
	A.~{Sprintson}, ``Non-adaptive quantitative group testing using irregular
	sparse graph codes,'' in \emph{Proc. 57th Annual Allerton Conference on
		Communication, Control, and Computing}, Sep. 2019, pp. 608--614.
	
	\bibitem{kautz}
	W.~Kautz and R.~Singleton, ``Nonrandom binary superimposed codes,''
	\emph{{IEEE} Trans. Inf. Theory}, vol.~10, no.~4, pp. 363--377, Oct. 1964.
	
\end{thebibliography}
\end{document}